\documentclass[preprint,prd,nofootinbib,tightenlines,groupedaddress,superscriptaddress,amsmath,amssymb]{revtex4}

\usepackage{graphicx}
\usepackage{bm}
\usepackage[hypertex]{hyperref}
\usepackage{color}

\begin{document}
\baselineskip=15pt \parskip=3pt

\vspace*{3em}


\title{Nonabelian Dark Matter with Resonant Annihilation}
\author{Cheng-Wei Chiang}
\affiliation{Department of Physics and Center for Mathematics and Theoretical Physics,
National Central University, Chungli 320, Taiwan\smallskip}
\affiliation{Institute of Physics, Academia Sinica, Taipei 115, Taiwan\smallskip}
\affiliation{Physics Division, National Center for Theoretical Sciences, Hsinchu 300, Taiwan\smallskip}
\author{Takaaki Nomura}
\affiliation{Department of Physics and Center for Mathematics and Theoretical Physics,
National Central University, Chungli 320, Taiwan\smallskip}
\affiliation{Department of Physics, National Cheng-Kung University, Tainan 701, Taiwan \smallskip}
\author{Jusak Tandean}
\affiliation{Department of Physics and Center for Theoretical Sciences,
National Taiwan University, Taipei 106, Taiwan\medskip}



\begin{abstract}
We construct a model based on an extra gauge symmetry, SU(2)$_X\times$U(1)$_{B-L}$, which can provide gauge bosons
to serve as weakly-interacting massive particle dark matter.
The stability of the dark matter is naturally guaranteed by a discrete $Z_2$ symmetry that is a subgroup of SU(2)$_X$.
The dark matter interacts with standard model fermions by exchanging gauge bosons which are linear combinations of  SU(2)$_X \times$U(1)$_{B-L}$ gauge bosons.
With the appropriate choice of representation for the new scalar multiplet whose vacuum expectation value spontaneously breaks the SU(2)$_X$ symmetry, the relation between the new gauge boson masses can naturally lead to resonant pair annihilation of the dark matter.
After exploring the parameter space of the new gauge couplings subject to constraints from collider data and the observed relic density, we use the results to evaluate the cross section of the dark matter scattering off nucleons and compare it with data from the latest direct detection experiments.
We find allowed parameter regions that can be probed by future direct searches for dark matter and LHC searches for new particles.
\end{abstract}
\maketitle

\section{Introduction}

The standard model (SM) of particle physics has been very successful in describing an enormous amount of experimental data at energies up to $\,{\cal O}(100)$ GeV.
There are, however, questions remaining that require physics beyond the minimal SM to address.
Among the outstanding issues are the explanations for the astronomical evidence of dark matter (DM) and for the numerous experimental indications of neutrino mass~\cite{pdg}.
It is then of great interest to explore a new physics scenario in which the DM and neutrino sectors are intimately connected.

Previously, we have considered a simple model which provides not only DM of the popular weakly-interacting massive particle (WIMP) type, but also a means to endow neutrinos with mass~\cite{CNT}.
The DM candidate belongs to a complex scalar singlet stabilized by a $Z_2$ symmetry that is not imposed in an {\it ad hoc} way, but instead emerges from an extra Abelian gauge group related to baryon number minus lepton number, U(1)$_{B-L}$, that is spontaneously broken by the nonzero vacuum expectation value (VEV) of a new scalar field, in the Krauss-Wilczek manner~\cite{Krauss:1988zc}.
Light neutrino masses are produced via the well-known seesaw mechanism~\cite{seesaw}, which is triggered with the involvement of the same new scalar field after the addition of right-handed neutrinos.
The DM relic density receives contributions mainly from diagrams mediated by the Higgs boson and also those mediated by the U(1)$_{B-L}$ gauge boson,~$Z'$.
It turns out that constraints from collider data and the observed relic density together imply that the $Z'$ mass has to be
in the resonance region of the $Z'$-mediated DM annihilation, namely about twice the DM mass.
Furthermore, results from DM direct detection experiments and Higgs data from the LHC favor the dominance of the $Z'$-exchange contributions to the relic density.
All this motivates us to look for a different possible scenario in which the resonance condition can be fulfilled naturally.

In this paper, we demonstrate that such a possibility can be realized in a model where the role of WIMP DM is played by massive gauge bosons associated with a nonabelian symmetry.
Although most of the WIMP DM candidates proposed in the literature are either fermionic or spinless, those with spin one have also been considered before~\cite{vectordm}.
Here we construct a model based on the gauge group \,G$_{\rm SM}\times$SU(2)$_X\times$U(1)$_{B-L}$,\, where G$_{\rm SM}$ refers to the SM group and the extra gauge symmetries offer gauge bosons which can act as WIMP candidates.
The stability of the DM is naturally maintained by a discrete $Z_2$ symmetry which is a subgroup of the new nonabelian gauge group,~SU(2)$_X$.
This $Z_2$ symmetry appears after the spontaneous breaking of SU(2)$_X$ by the nonzero VEV of a~new scalar multiplet, following the Krauss-Wilczek mechanism~\cite{Krauss:1988zc}.
Then the $Z_2$-odd gauge bosons associated with SU(2)$_X$ can serve as DM if they are lighter than other $Z_2$-odd particles in the model.
Since SM fermions are charged under U(1)$_{B-L}$, the DM can interact with SM fermions at tree level by exchanging gauge bosons which are obtained from the linear combinations of SU(2)$_X$ and U(1)$_{B-L}$ gauge fields.
Thus, the new gauge interactions are responsible for both the relic abundance and the DM interactions with nucleons.
Another interesting feature of the model is that, the DM being made up of SU(2)$_X$ gauge bosons, its mass is related to the masses of the mediating gauge bosons, implying that resonant pair annihilation can be naturally achieved by choosing suitable representations of the scalar fields involved in the breaking of the SU(2)$_X\times$U(1)$_{B-L}$ gauge symmetry and ensuring that their VEVs are sufficiently well separated.
What's more, the presence of the U(1)$_{B-L}$ gauge symmetry requires the introduction of right-handed neutrinos for gauge-anomaly cancellation, which in turn participate in the type-I seesaw mechanism to generate light neutrino masses~\cite{seesaw}, with the right-handed neutrino masses being connected to the U(1)$_{B-L}$ breaking scale.
This model turns out to have sufficient parameter space that is consistent with current collider, relic density, and DM direct search data.
Therefore, it can be probed further by ongoing or future DM direct detection experiments, and some of the new particles may be observable at the LHC with sufficient luminosities.

This paper is organized as follows.
The next section contains the details of our model which possesses WIMP DM composed of the gauge bosons of an extra nonabelian gauge symmetry.
We explain how the choices of the new particles and their quantum numbers can naturally translate into resonant annihilation of the DM.
In Section~\ref{const}, we examine constraints on the new gauge couplings from collider data.
In Section~\ref{relic}, we deal with the relic density of our DM candidates and extract the parameter values allowed by its observed value.
In Section~\ref{direct}, we use the results to predict the DM-nucleon scattering cross-section and compare it with current data from direct detection experiments.
In Section~\ref{collider}, we comment on the collider phenomenology of the new particles in our model.
We conclude in Section~\ref{summary} with the summary of our study and some more discussion.

\section{A model of dark massive gauge boson\label{model}}

\begin{table}[b] \vspace{1ex}
\begin{tabular}{|c||c|c|c|c|c|c|c|c|c|c|c|c|c|} \hline
$\vphantom{|_|^|}$ & $f_{\rm SM}^{}$ & $\nu_R^{}$ & $H$ & $S$ & $\phi_2^{}$ & $\phi_1^{}$ & $\phi_0^{}$ & $\phi_{-1}^{}$ & $\phi_{-2}^{}$ & $X$ & $X^\dagger$ & $C_3^{}$ & $E$
  \\ \hline \hline
\,SU(2)$_X\,\bigl[$U(1)$_{B-L}\bigr]\vphantom{|_|^|}$\, & \,1\,$[B-L]$\, & \,1\,$[-1]$\, & \,1\,[0]\, & \,1\,[2]\, & \,5\,[2]\, & \,5\,[2]\, & \,5\,[2]\, & \,5\,[2]\, & \,5\,[2]\, & \,3\,[0]\, & \,3\,[0]\, & \,3\,[0]\, & \,1\,[0]\, \\ \hline
$T_{3X}$ & $0$ & $0$ & $0$ & $0$ & $2$ & $1$ & $0$ & $-1$ & $-2$ & $1$ & $-1$ & $0$ & $0$ \\ \hline
$Z_2^X$ & $+$ & $+$ & $+$ & $+$ & $+$ &$-$ & $+$ & $-$ & $+$ & $-$ & $-$ & $+$ & $+$ \\ \hline
\end{tabular} \vspace{-1ex}
\caption{The charge assignments under SU(2)$_X^{}\times$U(1)$_{B-L}^{}$ and $Z_2^X$ parity
of the fermions, scalars and new gauge bosons in the model, with $f_{\rm SM}^{}$ referring to SM fermions, \,$X=(C_1-i C_2)/\sqrt{2}$,\, and $T_{3X}$ denoting the eigenvalue of the third generator of~SU(2)$_X$.\label{contents}} \vspace{-1ex}
\end{table}
%

Compared to the SM, the new model contains the additional gauge group SU(2)$_X\times$U(1)$_{B-L}$, where $X$ refers to the massive gauge boson that serves as the DM, whereas $B$ and $L$ stand for baryon and lepton numbers, respectively.
We denote the gauge fields associated with SU(2)$_X^{}$ and U(1)$_{B-L}^{}$ by $C_k^\mu$ and~$E^\mu$, respectively, \,$k=1,2,3$,\,
and their coupling constants $g_X^{}$ and $g_{B-L}^{}$.
The model also has new complex scalar fields $S$ and $\Phi_5$ as well as three extra fermions $\nu_{kR}^{}$, all of which are singlets under the SM gauge group, but carry nonzero U(1)$_{B-L}^{}$ charges.
Under SU(2)$_X$ transformations, $S$~is a~singlet, while $\Phi_5$ is a five-plet represented by the column matrix
\,$\Phi_5=\bigl(\phi_2^{},\phi_1^{},\phi_0^{},\phi_{-1}^{},\phi_{-2}^{}\bigr){}^{\rm T}$,\,
where $\phi_a^{}$ corresponds to the eigenvalue \,$T_{3X}=a$\, of the third generator of~SU(2)$_X$.
In Table~\ref{contents} we collect the SU(2)$_X\times$U(1)$_{B-L}$ quantum number assignments for the fermions, scalars, and new gauge bosons in the model, with $H$ being the usual scalar doublet.

The renormalizable Lagrangian for $S$ and $\Phi_5$, with $H$ included in the potential $\cal V$, is
\begin{eqnarray}
{\cal L} \,\,=\,\, \bigl({\cal D}^\mu S\bigr)^\dagger\, {\cal D}_\mu^{} S + \bigl({\cal D}^\mu \Phi_5^{}\bigr)^\dagger\, {\cal D}_\mu^{} \Phi_5^{} \,-\, {\cal V} ~,
\end{eqnarray}
where
\begin{eqnarray} & \displaystyle
{\cal D}^\mu S \,\,=\,\, \partial^\mu S+2i g_{B-L\,}^{}E^\mu S ~, \hspace{5ex}
{\cal D}^\mu\Phi_5^{} \,\,=\,\, \partial^\mu\Phi_5^{}+ig_{X\,}^{} C_k^{\mu\,}{\cal T}^{(5)}_k\Phi_5^{}+i g_{B-L\,}^{}E^{\mu}{\cal Q}_{B-L}^{(5)}\Phi_5^{} ~, &
\\  & \displaystyle
{\cal V} \,\,=\,\, -\mu_\Phi^2 \Phi_5^\dagger \Phi_5^{} + \bigl(\lambda_S^{} |S|^2 - \mu_S^2\bigr) |S|^2 + \bigl(\lambda_H^{} H^\dagger H - \mu_H^2\bigr)H^\dagger H
\;+\; (\mbox{other quartic terms}) ~. &
\end{eqnarray}
In $\,{\cal D}^\mu\Phi_5^{}$ above, summation over \,$k=1,2,3$\, is implicit, and ${\cal T}^{(5)}_k$ and ${\cal Q}_{B-L}^{(5)}$ are matrices for the generators of SU(2)$_X^{}$ and U(1)$_{B-L}^{}$, respectively, acting on $\Phi_5^{}$, where
\begin{eqnarray} & \displaystyle
{\cal T}_1^{(5)} \,\,=\,\, \frac{1}{2}
\begin{pmatrix}
0 & 2 & 0 & 0 & 0 \\
2 & 0 & \sqrt{6} & 0 & 0 \\
0 & \sqrt{6} & 0 & \sqrt{6} & 0 \\
0 & 0 & \sqrt{6} & 0 & 2 \\
0 & 0 & 0 & 2 & 0
\end{pmatrix} , \hspace{5ex}
{\cal T}_2^{(5)} \,\,=\,\, \frac{i}{2}
\begin{pmatrix}
0 & -2 & 0 & 0 & 0 \\
2 & 0 & -\sqrt{6} & 0 & 0 \\
0 & \sqrt{6} & 0 & -\sqrt{6} & 0 \\
0 & 0 & \sqrt{6} & 0 & -2 \\
0 & 0 & 0 & 2 & 0
\end{pmatrix} ,
& \nonumber \\ & \displaystyle
{\cal T}_3^{(5)} \,\,=\,\, {\rm diag}(2,1,0,-1,-2) ~, \hspace{5ex}
{\cal Q}_{B-L}^{(5)} \,\,=\,\, {\rm diag}(2,2,2,2,2) ~. &
\end{eqnarray}

In this paper, we consider the scenario in which the SU(2)$_X\times$U(1)$_{B-L}$ gauge symmetry is spontaneously broken according to
\begin{equation}
{\rm SU(2)}_X^{}\times{\rm U(1)}_{B-L}^{} \,\,\xrightarrow{\langle S\rangle}\,\, {\rm SU(2)}_X^{}\times Z_2^{B-L} \,\,\xrightarrow{\langle \Phi_5\rangle}\,\, Z_2^X \times Z_2^{B-L} ~,
\end{equation}
where \,$\langle S \rangle = v_S^{}/\sqrt{2}$\, and \,$\langle\Phi_5\rangle = \bigl(v_\Phi^{},0,0,0,0\bigr){}^{\rm T}/\sqrt{2}$\, are the VEVs of $S$ and $\Phi_5$, with \,$v_S^{}\gg v_\Phi^{}>0$.\,
Since \,$\langle\Phi_5\rangle\neq0$\, occurs via its \,$T_{3X}=2$\, component, \,$\langle\phi_2^{}\rangle\neq0$,\, the $Z_2^X$ symmetry emerges naturally as a subgroup of SU(2)$_X$ and the particles with even (odd) $T_{3X}$ values will be $Z_2^X$ even (odd), as Table~\ref{contents} shows.
On the other hand, $Z_2^{B-L}$ is the remnant of U(1)$_{B-L}^{}$ after \,$\langle S \rangle\neq0$,\, as discussed in~Ref.\,\cite{CNT}, but does not play a role in the stabilization of~$X$.
Thus, in this scenario the remaining $Z_2^X$ guarantees the stability of the lightest $Z_2^X$-odd particle(s), which can therefore act as DM.
Here we choose the gauge boson \,$X=(C_1-iC_2)/\sqrt{2}$\, and its conjugate $X^\dagger$ to be the DM, hence tacitly taking the $Z_2^X$-odd scalar bosons to be more massive than~$X$.
It is worth mentioning that we would arrive at the same results below if \,$\langle\Phi_5\rangle\neq0$\, through its \,$T_{3X}=-2$\, component instead.
As for $H$, its VEV is also nonvanishing and breaks the electroweak symmetry just as in the~SM.
We assume that the other parameters in the potential $\cal V$ are such that the vacuum has the above desired properties,
leaving a detailed analysis of $\cal V$ for future work.

After SU(2)$_X^{}\times$U(1)$_{B-L}^{}$ spontaneously breaks into \,$Z_2^X \times Z_2^{B-L}$,\, the new gauge bosons acquire in $\cal L$ the mass terms
\begin{eqnarray}
{\cal L}_{\rm m} &=& \bigl\langle\Phi_5^\dagger\bigr\rangle \left[g_{X\,}^{} C_{k\;}^\mu{\cal T}^{(5)}_k + g_{B-L}^{} E^{\mu}{\cal Q}_{B-L}^{(5)}\right]\left[g_{X\,}^{} C_{k'\mu}^{}{\cal T}^{(5)}_{k'} + g_{B-L\,}^{} E_{\mu}^{\;}{\cal Q}_{B-L}^{(5)}\right]
\bigl\langle\Phi_5^{}\bigr\rangle
\,+\,  4 g_{B-L\,}^2 E^2 \langle S\rangle^2
\nonumber \\
&=&  g_X^{2~} v_\Phi^{2\;} X_\mu^\dagger X^\mu \,+\, \frac{1}{2} \bigl( C_3^\mu ~~~ E^\mu \bigr)
\begin{pmatrix} 4 g_X^{2~} v_\Phi^2 \; & \; 4 g_X^{~~}g_{B-L\,}^{}v_\Phi^2 \\ 4 g_X^{~~}g_{B-L\,}^{}v_\Phi^2\; & \;4 g_{B-L}^2 \bigl(v_\Phi^2+v_S^2\bigr)_{\vphantom{|}} \end{pmatrix}
\begin{pmatrix} C_{3\mu}^{} \\ E_\mu^{} \end{pmatrix} .
\end{eqnarray}
From the last line, upon diagonalizing the 2$\times$2 matrix in the second term, we obtain the eigenvalues
\begin{eqnarray}
m_X^2 &=& g_X^{2~} v_\Phi^2 ~, \label{mX2} \\
 m_{Z_L,Z_H}^2 &=& 2 g_X^{2~} v_\Phi^2 + 2g_{B-L}^2 \bigl(v_\Phi^2+v_S^2\bigr) \mp 2 \sqrt{ \bigl[ g_X^{2~} v_\Phi^2-g_{B-L}^2\bigl(v_\Phi^2+v_S^2\bigr) \bigr]^2+4g_X^{2~} g_{B-L\,}^2 v_\Phi^4} ~,
 \end{eqnarray}
assuming that \,$m_{Z_L}<m_{Z_H}$\, for the mass eigenstates $Z_L$ and $Z_H$ which are given by
\begin{eqnarray}
&& \begin{pmatrix} Z_{L} \\ Z_{H} \end{pmatrix} \,=\,
 \begin{pmatrix} \cos \theta & ~~ \sin \theta \\ - \sin \theta & ~~ \cos \theta \end{pmatrix}_{\vphantom{|}}^{}
  \begin{pmatrix} C_3 \\ E \end{pmatrix} , \\
 \tan(2\theta) &=& \frac{2 g_X^{~~}g_{B-L\,}^{}R_v}{g_X^{2~}R_v - g_{B-L}^2 (1+R_v)} ~, \hspace{7ex}
 R_v \,\,=\,\, \frac{v_\Phi^2}{v_S^2}  ~. \label{Rv}
 \end{eqnarray}
In this study, we focus on the case in which \,$v_S^2\gg v_\Phi^2$\, and \,$g_X^{}\sim g_{B-L}^{}$,\, implying that
\begin{eqnarray}
 \label{Ap_angle}
 |\theta| &\simeq& \frac{g_X^{}}{g_{B-L}^{}}\,R_v^{} ~, \\
 \label{Ap_mZL}
 m_{Z_L}^2 &\simeq& 4 m_X^2 (1- R_v) ~, \\
 \label{Ap_mZH}
  m_{Z_H}^2 &\simeq& 4 m_X^2\, \frac{g_{B-L}^2}{g_X^{2~}R_v}(1+R_v) ~.
 \end{eqnarray}
Accordingly, with \,$R_v\ll1$,\, we obtain the mass relation
\begin{eqnarray} \label{resonant}
m_{Z_L}^{} \,\simeq\,\, 2 m_X^{} ~,
\end{eqnarray}
which naturally leads to resonant annihilation of the DM pair via the $Z_L$-mediated contribution.

It is worth noting that the five-plet $\Phi_5$ is the minimal choice of SU(2)$_X$ representation that can result in the resonant relation in~Eq.\,(\ref{resonant}).
In general, for an SU(2)$_X$ isospin value $T_X$ and its third component~$T_{3X}$, one would get
\,$m_X^2/m_{Z_L}^2 \simeq\bigl[T_X(T_X+1)-T_{3X}^2\bigr]/\bigl(2T_{3X}^2\bigr)$\, assuming small mixing angle~$\theta$,
in analogy to the $\rho$ parameter in the electroweak sector~\cite{pdg}.

The neutrino mass-generating sector is the same as that given in Ref.\,\cite{CNT}, the relevant Lagrangian having the form
\begin{equation}
{\cal L}_{m_\nu^{}} \,\,=\,\, i\lambda_{kl\,}^{}\bar\nu_{kR\,}^{}H^{\rm T}\tau_2^{}L_{lL}^{} - \mbox{$\frac{1}{2}$}\lambda_{kl\,}'\bar\nu_{kR\,}^{}(\nu_{lR})^{\rm c\,}S^\dagger \;+\; {\rm H.c.} \,,
\end{equation}
where summation over \,$k,l=1,2,3$\, is implicit, $\lambda_{kl}^{(\prime)}$ are free parameters, $\tau_2^{}$ is the second Pauli matrix, $L_{lL}$ represents a~lepton doublet, and the superscript c indicates charge conjugation.
The Dirac and Majorana mass matrices from these terms are
\,${\cal M}_D^{}=\lambda v_H^{}/\sqrt2$\, and \,${\cal M}_{\nu_R}=\lambda' v_S^{}/\sqrt2$,\,
respectively, where $v_H^{}$ is the VEV of $H$.
Hence $v_S^{}$ sets the mass scale of the right-handed neutrinos,~$\nu_{kR}^{}$.
In our examples later on, we will see what values of $v_S^{}$ are compatible with the observed relic density and collider data.

Since $X$ is our chosen candidate for DM and interacts with SM fermions by exchanging the $Z_{L,H}$ bosons at tree level, in the following two sections we evaluate the new gauge couplings subject to collider and relic density data.
Subsequently, we use the allowed values of the couplings to predict the cross section of the DM-nucleon scattering and compare it with the existing results of DM direct detection experiments.

\section{Constraints from collider experiments\label{const}}

The gauge bosons $Z_L$ and $Z_H$ interact with SM fermions at tree level with coupling constants \,$g_{B-L}^{}\sin\theta$\, and \,$g_{B-L}^{}\cos\theta$,\, respectively, according to the Feynman rules listed in Appendix~\ref{FR}.
It follows that measurements on processes mediated by $Z_L$ and $Z_H$ can offer constraints on these couplings.
Significant restrictions may be available from the data on $e^+e^-$ and hadron collisions into fermion pairs, which we treat in this section.

We first look at the constraints from \,$e^+ e^-\to f\bar f$\, scattering.
In this work we assume that mixing between the $Z$ boson and $Z_{L,H}$ is negligible, but we will comment on the impact of kinetic mixing between them later on and discuss it further in Appendix~\ref{KM}.
In the absence of the mixing, the new gauge couplings have no effects on the $Z$-pole observables at leading order.
On the other hand, the measurements of \,$e^+ e^-\to f\bar f$\, at LEP\,II with center-of-mass energies from 130 to 207~GeV are relevant~\cite{Alcaraz:2006mx}.
We employ the data on the cross section and forward-backward asymmetry for \,$f=\mu,\tau$\, and
on the cross section for \,$f=\rm quark$.\,
To evaluate the limits on the new couplings, we include both the $Z_L$ and $Z_H$ contributions to the scattering amplitude, their couplings and masses satisfying the relations in Eqs.~(\ref{Ap_angle})-(\ref{Ap_mZH}).
Although $Z_H$ is much heavier than $Z_L$, the fermionic couplings of the latter can be much smaller than those of the former to compensate for the suppression of the $Z_H$ contribution to the amplitude due to its bigger mass.
In the examples presented below, the $Z_H$ contributions to \,$e^+ e^-\to f\bar f$\, turn out to dominate the $Z_L$ ones.

\begin{figure}[b]
\includegraphics[width=87mm]{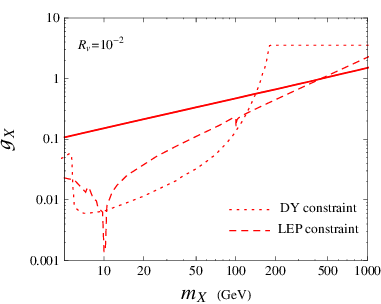}\includegraphics[width=87mm]{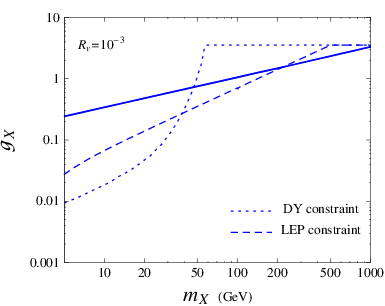}\vspace{-1ex}
\caption{Upper limits on $g_X^{}$ versus $m_X^{}$ from LEP~II and LHC data on \,$e^+e^-\to f\bar f$\, and Drell-Yan scattering, respectively, for \,$R_v=10^{-2}$ (left) and $10^{-3}$ (right)\, under the assumption that \,$g_X^{}=g_{B-L}^{}$,\, compared to the corresponding values of $g_X^{}$ (solid curves) consistent with the observed relic density.
The horizontal, straight portions of the dashed and dotted curves correspond to the perturbativity condition, \,$g_X^{}<\sqrt{4\pi}$.\label{gXmX}}
\end{figure}

For definiteness and simplicity, hereafter we set \,$g_X^{}=g_{B-L}^{}\ge0$.\,
Adopting the 90\% confidence-level (CL) ranges of the LEP\,II measurements~\cite{Alcaraz:2006mx} and using the formulas given
in~Ref.\,\cite{Zprime}, but with $s$-dependent $Z$ and $Z_{L,H}$ widths~\cite{Alcaraz:2006mx}, we then scan the $m_X^{}$ and $R_v$ space.
To illustrate the results, we display in Figure~\ref{gXmX} the upper limits on $g_X^{}$ versus $m_X^{}$ for \,$R_v=10^{-2}$ (red dashed curve) and $10^{-3}$ (blue dashed curve)\, on the left and right sides, respectively.
The horizontal, straight portions of the curves correspond to the perturbativity requirement, \,$g_X^{}<\sqrt{4\pi}$.\,
%

The most recent data from the LHC on the cross-section of the Drell-Yan (DY) process in proton-proton collisions at \,$\sqrt s=7$\,TeV\,  with $4.5{\rm\,fb}^{-1}$ of integrated luminosity have revealed no discrepancy from the SM expectations and therefore no evidence of $Z_{L,H}$ bosons~\cite{dy@lhc}.
Consequently, we follow the same analysis as in Refs.~\cite{CW, CNT} to derive upper bounds on the coupling constants using the SM cross-section.
In the present case, we can consider the $Z_L$ and $Z_H$ contributions separately because we focus on events with dilepton invariant mass around $m_{Z_{L}}$ or $m_{Z_{H}}$ where their effects are of different orders for small $R_v$.
Thus respective constraints are obtained for the pairs $\bigl(m_{Z_L}, g_{B-L}^{}\sin \theta\bigr)$ and $\bigl(m_{Z_H},g_{B-L}^{}\cos\theta\bigr)$.
To estimate the DY cross-section numerically, we utilize the CalcHEP package~\cite{Ref:CalcHEP} by incorporating the new particles and Feynman rules of our model.
Then we apply the one-bin log likelihood \,$LL=2[N \ln(N/\nu) + \nu - N]$,\,  where $N$ ($\nu$) is the number of events predicted by
the SM (SM plus the $Z_L$ or $Z_H$ boson) in the $\ell^+\ell^-$ invariant mass window of $\pm 20 \%$ around the expected $Z_L$ or $Z_H$ mass, with \,$\sqrt s=7$\,TeV\, and $4.5{\rm\,fb}^{-1}$ of luminosity.
The upper limit on the cross-section is obtained from the solved value of $\nu$ for each $Z_L$ or $Z_H$ mass,
after adopting \,$LL=2.7$\, which corresponds to the $90 \%$ CL.
We find that the $Z_H$ contribution to the DY process yields a stricter bound on $g_{B-L}^{}$ as a function of $m_{Z_H}$, as the $Z_L$ contribution is strongly suppressed by the small $|\theta|$.
We show the resulting upper-limits on $g_X^{}\bigl(=g_{B-L}^{}\bigr)$ in Figure~\ref{gXmX}, where $m_X^{}$ is related to $m_{Z_H}$ by Eq.\,(\ref{Ap_mZH}), for \,$R_v=10^{-2}$ (red dotted curve) and $10^{-3}$ (blue dotted curve)\, on the left and right, respectively.
We notice that the limit in the \,$R_v=10^{-2}$\, case becomes large at \,$m_X^{}\sim 5$\,GeV\, corresponding to \,$m_{Z_H}\sim m_Z^{}$\, where the SM background is large.
%

\section{Resonant dark matter annihilation and relic density \label{relic}}

Now we estimate the relic density of the DM particle, $X$, in order to search for the model parameter space consistent with the observed relic density.
The thermal relic abundance is found by solving the Boltzmann equation which describes
the number density of the DM.
We employ the approximate solution to the Boltzmann equation for the present-day relic density $\Omega$, given by~\cite{Kolb:1990vq}\footnote{For a more accurate approximation, see~\cite{Steigman:2012nb}.}
\begin{eqnarray} \label{omega} & \displaystyle
\Omega h^2 \,\,=\,\, \frac{1.07\times10^9}{\sqrt{g_*^{}}\;m_{\rm Pl}^{}\,J\,\rm\,GeV} ~, \hspace{5ex}
J  \,\,=\,\, \raisebox{0.7ex}{\footnotesize$\displaystyle\int_{x_f}^\infty$} dx\;
\frac{\langle\sigma v\rangle}{x^2} ~,
& \nonumber \\ & \displaystyle
x_f^{} \,\,=\,\, \ln\frac{0.038\,g\,m_X^{}\,m_{\rm Pl}^{}\,\langle\sigma v\rangle}{\sqrt{g_*^{}x_f^{}}} ~, &
\end{eqnarray}
where $h$ denotes the Hubble constant in units of 100\,km/s/Mpc, $g_*^{}$ is the number of relativistic degrees of freedom below the freeze-out temperature~\,$T_f^{}=m_X^{}/x_f^{}$,\,  \,$m_{\rm Pl}=1.22 \times 10^{19}$\, GeV is the Planck mass, \,$g=3$\, to account for $X$ having spin-1, and $\langle\sigma v\rangle$ is the thermal average of the DM annihilation cross-section.\,
More explicitly~\cite{TA},
\begin{eqnarray} \label{sv}
\langle\sigma v\rangle \,\,=\,\, \frac{x}{8 m_X^{5\;}K_2^2(x)}
\int_{4m_X^2}^\infty ds\;\sqrt s\,\bigl(s-4m_X^2\bigr)\;
K_1^{}\bigl(\sqrt s\,x/m_X^{}\bigr)\,\sigma_{\rm ann}^{} ~,
\end{eqnarray}
where $K_i$ is the modified Bessel function of the second kind of order $i$ and $\sigma_{\rm ann}^{}$ represents the cross section of $X^\dagger X$ annihilation into all possible final states.

Under the assumptions made in Section \ref{model}, we find that the main contributions to $\sigma_{\rm ann}^{}$ come from
the $s$-channel transitions \,$X^\dagger X\to Z_L^*\to f_{\rm SM}^{}\bar f_{\rm SM}^{}$.\,
Although $Z_H$-mediated diagrams also contribute, in this case they can be neglected because of the suppression due to \,$m_{Z_H}\gg m_{Z_L}$\, and their lack of the resonance enhancement of the $Z_L$-mediated diagrams in the nonrelativistic region \,$\sqrt s\sim 2m_X^{}$\, due to \,$m_{Z_L}\simeq 2m_X^{}$.\,
Thus, with the Feynman rules in Appendix~\ref{FR}, we arrive at
\begin{eqnarray} \label{sann}
\sigma_{\rm ann}^{} &=&
\frac{g_X^{2~} g_{B-L}^2 \cos^2\!\theta \sin^2\!\theta}{432\pi}\sum_f
\frac{\sqrt{\bigl(s-4m_X^2\bigr)\bigl(s-4m_f^2\bigr)}}{m_X^4\,s}\;
\frac{s^2+20\,m_X^2\,s+12\,m_X^4}
{\bigl(s-m_{Z_L}^2\mbox{$\bigr)^2$}+\Gamma_{Z_L}^2m_{Z_L}^2}
\nonumber \\ && \hspace{24ex} \times\;
\Bigl[ \bigl(s+2m_f^2\bigr) \bigl|\hat V_f^{Z_L}\bigr|^2 +
\bigl(s-4m_f^2\bigr) \bigl|\hat A_f^{Z_L}\bigr|^2 \Bigr] N_{\rm c}^f ~,
\end{eqnarray}
where the sum is over all fermions with masses \,$m_f^{}<m_X^{}$\, and color factors $N_{\rm c}^f$, the couplings $\hat{V}_f^{Z_L}$ and $\hat{A}_f^{Z_L}$ are given in Eq.~(\ref{Zff}), and $\Gamma_{Z_L}$ is the width of $Z_L$.
Now, since \,$m_{Z_L}^2=4 m_X^2(1- R_v)$\, and \,$s\ge 4m_X^2$\, according to Eqs.~(\ref{Ap_mZL}) and~(\ref{sv}), respectively, in the denominator of $\sigma_{\rm ann}^{}$ above we have~\,$\bigl(s-m_{Z_L}^2\bigr){}^2\ge 16 m_X^4 R_v^2$.\,
From the collider bounds on \,$g_{B-L}^{}=g_X^{}$\, derived in the previous section, we find that for the mass range of interest \,$16 m_X^4 R_v^2\gg\Gamma_{Z_L}^2m_{Z_L}^2$.\,
Consequently, the $\Gamma_{Z_L}$ term can be neglected in the calculation of Eq.\,(\ref{sv}).

With Eqs. (\ref{omega})-(\ref{sann}), we can extract the $\big(g_X^{},m_X^{}\bigr)$ regions compatible with the observed~$\Omega$.
Its most recent value has been determined by the Planck Collaboration from the Planck measurement and other data to be \,$\Omega h^2=0.1187\pm0.0017$\,~\cite{planck}.
Accordingly, we require the relic density of $X$ to satisfy the 90\% CL (confidence level) range of its experimental value,~\,$0.1159\le\Omega h^2\le0.1215$.\,
As mentioned in the preceding section, for simplicity we take \,$g_X^{}=g_{B-L}^{}$,\, implying that \,$|\theta|\simeq R_v$.\,
The plots in Figure~\ref{gXmX} display the resulting $g_X^{}$ values allowed by the relic data for \,$R_v=10^{-2}$ (red solid curve) and $10^{-3}$ (blue solid curve)\, on the left and right panels, respectively.
One can see that, although the $f\bar f Z_L$ couplings are suppressed by the small mixing angle, \,$|\theta|\ll 1$,\, the observed relic density can be reproduced with moderate-sized couplings $g_{B-L}^{}=g_X^{}={\cal O}(0.1)$-${\cal O}(1)$\, over \,$m_X^{}\leq 1000$\,GeV\, due to the resonance enhancement.
This can be partly understood from the fact that in the resonance region the denominator of $\sigma_{\rm ann}$ is dominated by the term \,$\bigl(4m_X^2-m_{Z_L}^2\bigr){}^2\propto R_v^2$\, which approximately cancels the $R_v^2$ factor in the numerator.

In Figure~\ref{gXmX}, we can also compare the coupling ranges that fulfill the requirements from both the collider and relic density data.
Evidently, the constraints from LEP\,II data restrict the allowed masses to \,$m_X^{}\,\mbox{\small$\gtrsim$\;}400\,(220)$~GeV\, with couplings of ${\cal O}(1)$ for \,$R_v=10^{-2}\,\bigl(10^{-3}\bigr)$.\,
The cases with \,$R_v \lesssim 10^{-4}$\, and \,$m_X^{}\leq 1000$\,GeV\, are excluded by the LEP\,II constraints.

\begin{figure}[t]
\includegraphics[width=123mm]{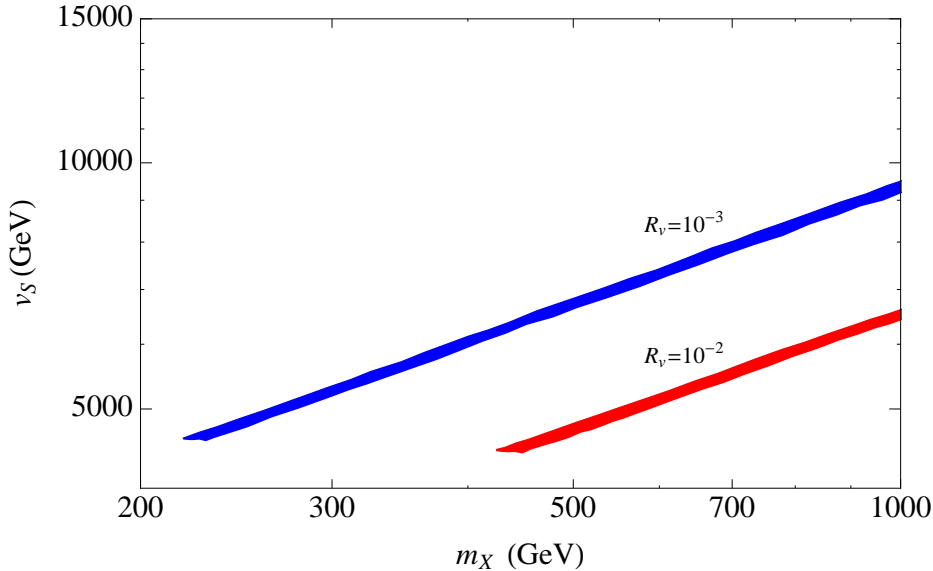}
\caption{Values of $v_S^{}$ versus $m_X^{}$ satisfying the requirements from both the collider and relic density data and corresponding to the allowed $g_X^{}$ regions in Figure~\ref{gXmX}.\label{mXvS}}
\end{figure}

Since we have the relation \,$m_X^{}=g_X^{}\,v_S^{}\sqrt{R_v}$\, from Eqs. (\ref{mX2}) and (\ref{Rv}), it is interesting to explore the values of $v_S^{}$ subject to the same experimental requirements.
We illustrate this in Figure~\ref{mXvS} obtained with the allowed $g_X^{}$ ranges in Figure~\ref{gXmX}.
Hence $v_S^{}$ should be between about 5 and 10~TeV in order to satisfy both the collider and relic data.
This suggests that our model is compatible with the TeV-scale type-I seesaw scenario.

\section{Direct detection of dark matter \label{direct}}

The direct detection of $X$ relies on its scattering off a nucleon~$N$ elastically, \,$XN\to XN$,\,
which proceeds from $Z_{L, H}$ exchanges in the $t$ channel.
Since \,$m_{Z_H}\gg m_{Z_L}$, the $Z_H$ contribution can be neglected.
It follows that in the nonrelativistic limit the cross section of \,$XN\to XN$\, is
\begin{equation}
\sigma_{\rm el}^N \,\,=\,\,
\frac{g_X^{2~} g_{B-L}^2 \cos^2\!\theta \sin^2\!\theta\, \mu_{XN}^2}{\pi m_{Z_L}^4}
\,\,\simeq\,\, \frac{g_X^{4\,} R_v^2\,\mu_{XN}^2}{16\pi\,m_X^4} ~,
\end{equation}
where \,$\mu_{XN}^{}=m_X^{}m_N^{}/\bigl(m_X^{}+m_N^{}\bigr)$\, and we have made use of
\,$\langle N|\bar u\gamma^\alpha u+\bar d\gamma^\alpha d|N\rangle=3\bar N\gamma^\alpha N$\,~\cite{NucleonMatrixElement}, the other quarks having vanishing contributions.
This indicates that $\sigma_{\rm el}^N$ is not sensitive to $g_{B-L}^{}$ for fixed \,$R_v\ll1$\,.

\begin{figure}[ht]
\includegraphics[width=111mm]{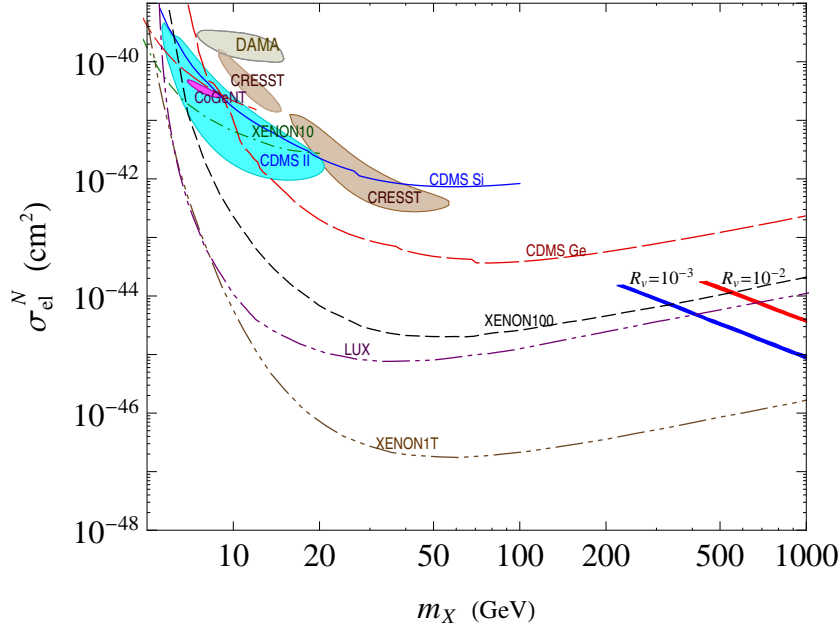}
\caption{Cross-section $\sigma_{\rm el}^N$ of \,$XN\to XN$\, scattering corresponding to the allowed parameter regions in Figure~\ref{gXmX}.
The predicted cross-sections are compared to $90\%$\,CL upper-limits from XENON10 (green dashed-dotted curve)~\cite{XENON10,Frandsen:2013cna}, XENON100 (black short-dashed curve)~\cite{XENON100}, CDMS Ge (red long-dashed curves)~\cite{CDMS-Ge}, CDMS Si (blue solid curve)~\cite{CDMS-Si}, and LUX (purple dashed-double-dotted curve)~\cite{Akerib:2013tjd}.
The prediction is also compared to the 90\%\,CL (magenta) signal region suggested by CoGeNT~\cite{cogent},
a gray patch compatible with the DAMA modulation signal at the 3$\sigma$ level~\cite{DAMA},
two 2$\sigma$-confidence (light brown) areas representing CRESST-II data~\cite{CRESST}, and a cyan area for a possible signal at 90\%\,CL from CDMS II~\cite{CDMSII}.
Also plotted is the XENON1T projected sensitivity (brown dashed-triple-dotted curve)~\cite{xenon1t}.
\label{Nscattering}}
\end{figure}

In Figure~\ref{Nscattering} we plot $\sigma_{\rm el}^N$ as a function of $m_X$ for the allowed parameter regions in Figure~\ref{gXmX}, the red and blue strips belonging to the \,$R_v=10^{-2}$ and $10^{-3}$\, cases, respectively.
Also shown are the recent data from  DM direct searches.
Clearly, much of the $\sigma_{\rm el}^N$ prediction still escapes the existing constraints, including the strictest ones from XENON100~\cite{XENON100} and LUX~\cite{Akerib:2013tjd}, but it will be probed more stringently by future direct searches such as XENON1T~\cite{xenon1t}.

Before moving on, we would like to make a few remarks regarding the potential implications
of mixing between the SM and extra gauge bosons in our model.
Since none of the scalar fields in the theory carries both the electroweak and new quantum
numbers, there is no mass mixing between the SM and new gauge bosons.
In contrast, as discussed in Appendix~\ref{KM}, kinetic mixing between the U(1)$_Y$
and U(1)$_{B-L}$ gauge bosons can occur both at tree and loop levels.
We find that the impact of this mixing is not significant on the results above for the allowed
values of the new gauge couplings and $Z_L$ mass.
Especially, the relation \,$m_{Z_L}\simeq 2m_X^{}$\, is unaffected.
We further find that, although the $Z_H$ mass is sensitive to the kinetic mixing,
being enhanced by it, the effect can be minimized if the mixing parameter has a magnitude
below~0.5.
Our rough estimate of the relevant loop diagram in Appendix~\ref{KM} suggests that mixing
size of order 0.5 is not atypical.
Lastly, since the $X$ annihilation and $X$-nucleon scattering processes are dominated by
the $Z_L$ contributions, the increased $m_{Z_H}$ would not be important for them.
It follows that it is reasonable to neglect the impact of the kinetic mixing.

\section{Comments on collider phenomenology\label{collider}}

In this section, we briefly discuss how the extra scalar and gauge bosons in our model may be produced and detected at the LHC.
The new scalar bosons coming from $\Phi_5$ and $S$ comprise twelve degrees of freedom in total.  Four of them are ``eaten'' by the new gauge bosons, making them massive.
The remaining extra scalar bosons can be expressed as $\phi_i^{\rm even}$ and $\phi_j^{\rm odd}$,
which are linear combinations of $Z_2^X$-even and -odd particles, respectively.
Since two of the new massive gauge bosons are $Z_2^X$ even and the other two $Z_2^X$ odd,
there are six $\phi^{\rm even}$'s and two $\phi^{\rm odd}$'s which are physical.
In this study we do not specify the new scalars' mass spectrum, but one could obtain it by doing a detailed analysis of the scalar potential.

Taking into account the $Z_2^X$ parities of the new particles, we find the decay patterns
\begin{eqnarray}
\label{decayZL}
& Z_{L} \,\,\to\,\, f_{\rm SM\,}^{} \bar{f}_{\rm SM}^{} ~, \\
\label{decayZH}
& Z_{H} \,\,\to\,\, f_{\rm SM\,}^{} \bar{f}_{\rm SM}^{}, \ XX^\dagger, \ \phi^{\rm even}_i \phi^{\rm even}_j, \ \phi^{\rm odd}_i \phi^{\rm odd}_j ~, \\
\label{decayPeven}
& \phi^{\rm even}_i \,\,\to\,\, Z_{L,H}^{} Z_{L,H}^{}, \ XX^\dagger, \ X^{(\dagger)} \phi^{\rm odd}_i, \ Z_{L, H\,}^{} \phi^{\rm even}_i,
 \  \phi^{\rm even}_j \phi^{\rm even}_k, \ \phi^{\rm odd}_j \phi^{\rm odd}_k ~, \\
\label{decayPodd}
& \phi^{\rm odd}_i \,\,\to\,\, Z_{L,H}^{} X, \  \phi^{\rm even}_j X, \ \phi^{\rm even}_j \phi^{\rm odd}_k ~,
\end{eqnarray}
where the particles on the right-hand sides may be off-shell depending on the masses involved.
Throughout we have assumed that $X$ is lighter than new scalar bosons, and so $Z_L$ decays mostly to SM fermions.
Since the couplings of $Z_{L,H}$ to the fermions are proportional to their $B-L$ numbers, $Z_{L,H}$ tend to decay
into leptons rather than quarks, as the decay rates of $Z_{L,H}$ into a~charged lepton pair and into a quark-antiquark pair, with relatively negligible masses, are related by \,$\Gamma_{Z_{L,H}\to\ell^+\ell^-}:\Gamma_{Z_{L,H}\to q\bar q}\simeq 1:3 (1/3)^2 = 3:1$.\,
The decay branching fractions of the scalar bosons depend on their mass spectrum and couplings in the potential.

The DM pair, $XX^\dagger$, can be produced through $Z_{L,H}$ exchanges in the $s$-channel according to
\begin{equation}
pp \,\,\to\,\,  Z_{L,H\;}^{(*)} +\; [{\rm jet(s),\,photon(s),\, etc.}] \,\,\to\,\,
XX^\dagger\, +\; [{\rm jet(s),\,photon(s),\, etc.}] ~,
\end{equation}
where we need particles other than $X X^\dagger$, such as jets ($j$'s) or photons ($\gamma$'s), for tagging.
Their production signals are therefore jet(s) plus missing energy, photon(s) plus missing energy, etc.
We note that the $Z_L$-mediated contributions dominate these processes because $Z_H$ is much heavier than~$Z_L$.

We now estimate the cross-sections of \,$pp\to X X^\dagger j$\, and \,$pp\to XX^\dagger\gamma$\, using the CalcHEP code package~\cite{Ref:CalcHEP} incorporating the new Feynman rules in the model file.
The cross sections are computed for the Tevatron and the LHC with different colliding energies employing two parameter sets taken from Figure~\ref{gXmX}.
The results are listed in Table~\ref{cross_section}.
These cross sections are small compared to current search limits due to the suppression by the small $|\theta|$.
For example, it is estimated that the upper limit of the cross section for jets plus missing energy in the squark-gluino-neutralino scenario of the minimal supersymmetric standard model is $\cal O$(1)\,fb for the LHC at \,$\sqrt s=7$\,TeV\, with $4.7\,\rm{fb}^{-1}$ of luminosity~\cite{susylimit}, which is larger than our cross sections.
However, at the LHC the cross-section of \,$pp\to XX^\dagger j$\, can reach \,{\small$\sim$}\,0.1\,fb\, for \,$\sqrt s=8$\,TeV\, and \,{\small$\sim$}\,0.5\,fb\, for \,$\sqrt s=14$\,TeV,\, which are potentially testable with the appropriate amount of luminosity.
The DM can also be produced singly in association with $\phi^{\rm odd}_i$, such as in
\begin{equation} \label{XPodd}
pp \,\,\to\,\, Z_{L,H}^{(*)} \,+\, \cdots \,\,\to\,\, X^{(\dagger)} \phi^{\rm odd}_i \,+\, \cdots ~.
\end{equation}
Since $\phi^{\rm odd}$ would decay according to Eq.~(\ref{decayPodd}), the specific signal would be charged leptons plus missing energy:
 \,$X^{(\dagger)} \phi^{\rm odd}_i \to X X^\dagger Z_{L} \rightarrow XX^\dagger \ell^+ \ell^-$\, where $\ell$ is the electron or muon.
The cross section of this channel is expected to be of similar order as that of $XX^\dagger$ production since the coupling constants involved are the same.

\begin{table}[b] \vspace{2ex}
\begin{tabular}{|l|c|c|c|c|c|} \hline
& \,$m_X^{}$ (GeV)\, & ~$g_X^{}$~ & $|\theta|$ & \, $\sigma_{XX^\dagger j_{\vphantom{|}}}^{}$ (fb) \, & \, $\sigma_{XX^\dagger\gamma}^{}$ (fb) \, \\ \hline
Tevatron & 300 & 1.8 & 0.001 & $6.5 \times 10^{-3}$  & $1.7 \times 10^{-4}$  \\
                & 600 & 1.2 & 0.01 & $6.0 \times 10^{-5}$ & $1.8 \times 10^{-6}$ \\ \hline
LHC 7 TeV & 300 & 1.8 & \,0.001\, & $8.7\times 10^{-2}$ & $1.6 \times 10^{-3}$ \\
                     & 600 & 1.2 & 0.01 & $5.1 \times 10^{-2}$ & $1.1 \times 10^{-3}$ \\ \hline
LHC 8 TeV & 300 & 1.8 & 0.001 & $8.6 \times 10^{-2}$ & $2.3 \times 10^{-3}$ \\
                     & 600 & 1.2 & 0.01 & 0.12 & $1.9 \times 10^{-3}$ \\ \hline
LHC 14 TeV & 300 & 1.8 & 0.001 & 0.46 & $8.2 \times 10^{-3}$ \\
                       & 600 & 1.2 & 0.01 & 0.51 & $1.0\times 10^{-2}$ \\ \hline
\end{tabular}
\caption{Estimated cross-sections of \,$pp \ (p \bar{p}) \rightarrow XX^\dagger j$\, and
\,$pp \ (p \bar{p}) \rightarrow XX^\dagger\gamma$\, for two parameter sets taken from Figure~\ref{gXmX} .
\label{cross_section}}
\end{table}

The scalar bosons $\phi^{\rm even}_i$ and $\phi^{\rm odd}_i$ can be produced through $Z_{L,H}$ exchanges in the $s$ channel,
\begin{eqnarray}
& pp \,\,\to\,\, Z_{L,H}^* \,\,\to\,\, Z_{L,H\,} \phi^{\rm even}_i ~, \\
& pp \,\,\to\,\, Z_{L,H}^{(*)} \,\,\to\,\, \phi^{\rm odd}_i \phi^{\rm odd}_j ~, \\
& pp \,\,\to\,\, Z_{L,H}^{(*)} \,\,\to\,\, \phi^{\rm even}_i \phi^{\rm even}_j ~,
\end{eqnarray}
as well as the channels in Eq.~(\ref{XPodd}).
According to the decay patterns in Eqs.~(\ref{decayPeven}) and~(\ref{decayPodd}), the signals of $\phi^{\rm even}_i$ and $\phi^{\rm odd}_i$ would be two pairs of charged leptons and charged leptons plus missing energy, respectively.
The production cross-sections of the $Z_{L}$-exchange processes are suppressed by $|\theta|^2$ because the $Z_L f\bar{f}$ coupling is proportional to $\sin \theta$, whereas the production cross-sections of $Z_H$-exchange processes are suppressed by the heavy $Z_H$ mass.
Thus high luminosities will be required to get a sufficient number of signal events.
To carry out a detailed analysis, one would need to specify the self-couplings and thus the mass spectrum of scalar bosons.  We leave such a~study for future work.
%

\section{Summary and discussion\label{summary}}

In this paper we have constructed a model possessing an extra gauge symmetry,  SU(2)$_X \times$U(1)$_{B-L}$,
 which offers a massive gauge boson, $X$, associated with SU(2)$_X$ playing the role of WIMP~DM.
The new gauge bosons become massive after SU(2)$_X$ and U(1)$_{B-L}$ are spontaneously broken by scalar fields $\Phi_5$ and $S$ developing nonzero VEVs of $v_\Phi^{}$ and $v_S^{}$, respectively, with \,$v_S^{}\gg v_\Phi^{}$.\,
The stability of the DM candidate is guaranteed by a residual $Z_2$ symmetry that is a subgroup of SU(2)$_X$.
At tree level, the dark gauge boson $X$ can interact with SM fermions by exchanging new gauge bosons $Z_{L,H}$ which arise from linear combinations of SU(2)$_X$ and U(1)$_{B-L}$ gauge fields.
The  $X X^\dagger$ pair annihilates into SM fermions by exchanging $Z_{L,H}$ in the $s$ channel.
Since the DM is a gauge boson, its mass can be related to the masses of other gauge bosons in the model.
The relation \,$m_{Z_L} \simeq 2m_X$\, emerges when SU(2)$_X$ is broken by the VEV of the SU(2)$_X$ scalar five-plet $\Phi_5$,
 naturally leading to resonant pair annihilation of $XX^\dagger$ via $Z_L$ exchange.
This model also supplies light neutrino masses with the aid of right-handed neutrinos whose mass terms are generated
when U(1)$_{B-L}$ is spontaneously broken by the VEV of $S$, which activates the type-I seesaw mechanism.

We considered the values of the new gauge couplings $g_X^{}$ and $g_{B-L}^{}$ in the case where they are equal
subject to constraints from collider data and the observed relic density.
Assuming that \,$R_v=v_\Phi^2/v_S^2\ll1$\, and \,$m_X \le1$\,TeV,\, we obtain no parameter space which survives these experimental restrictions for~\,$R_v\lesssim 10^{-4}$,\, but the \,$m_X^{}\,\mbox{\small$\gtrsim$\;}400\,(220)$~GeV\, region with ${\cal O}(1)$ couplings is still allowed for \,$R_v=10^{-2}\,\bigl(10^{-3}\bigr)$.\,
We also find that the corresponding values of $v_S^{}$ are between 5 and 10~TeV, implying that our model prefers the TeV-scale type-I seesaw scenario.
Subsequently, we explored the $X$-nucleon elastic scattering cross-section, $\sigma_{\rm el}^N$, for the surviving parameter regions and compared it with data from the latest DM direct detection experiments.
The resulting $\sigma_{\rm el}^N$ largely evades the most recent XENON100 and LUX limits and will be tested more strictly by future DM direct searches such as XENON1T.

Finally, we schematically discussed some of the phenomenology of the new particles at the LHC.
The DM particle can be produced as \,$XX^{\dagger}$ or $X^{(\dagger)} \phi^{\rm odd}_i$,\, where $\phi^{odd}_i$ is a $Z_2^X$-odd scalar boson.
The signals of these production processes would be missing energy plus jets/photons and
missing energy plus charged lepton pairs, respectively.
The new scalar bosons can also be produced as \,$\phi^{\rm even(odd)}\phi^{\rm even(odd)}$ or $\phi^{\rm odd}X\,(\phi^{\rm even} Z_{L,H})$,\, and the signals for $\phi^{\rm even}$ and $\phi^{\rm odd}$ would be two pairs of charged leptons and missing energy plus charged leptons, respectively.
Their production cross-sections tend to be suppressed due to the small $|\theta|$ value and/or the heavy $Z_H$ mass.
Nevertheless, our model would be testable with an appropriate luminosity in the future.
To perform a more detailed study would require specifying the self-couplings  in the scalar potential and thus the mass spectrum of the scalar bosons.
%

\acknowledgments

This research was supported in part by the National Science Council of Taiwan, R.O.C., under Grants No. NSC-100-2628-M-008-003-MY4 and No. NSC-100-2811-M-002-090 and by the MOE Academic Excellence Program under Grant No. 102R891505.

\appendix

\section{Feynman rules for new gauge interactions \label{FR}}

The couplings of the fermion $f$ in the model with the $Z_{L,H}$ bosons come from the U(1)$_{B-L}$ gauge interaction of $f$ described by
 \begin{equation}
{\cal L}' \,\,=\,\, -g_{B-L\,}^{}\bar f\gamma^\mu \Bigl(\hat{L}_{f}^{E}P_L^{}+\hat{R}_{f}^{E} P_R^{}\Bigr) f\, E_\mu^{} ~,
 \end{equation}
where \,$P_{L,R}=\frac{1}{2}(1\mp \gamma_5)$\, and \,$\hat{L}_{f}^{E\,}\bigl(\hat{R}_{f}^{E}\bigr)$\, is the \,$B-L$\, value for $f_{L(R)}$.
Since \,$E=Z_{L}\sin\theta+Z_{H}\cos\theta$,\, this leads to the Feynman rules
\begin{eqnarray}
\label{Zff}
& \bar{f} f Z_L^\mu : \; -i g_{B-L}^{}\,\sin \theta \left( \hat{V}_{f}^{E} + \hat{A}_{f}^{E} \gamma_5 \right) \gamma^\mu ~, \\
& \bar{f} f Z_H^\mu : \; -i g_{B-L}^{}\,\cos \theta \left( \hat{V}_{f}^{E} + \hat{A}_{f}^{E} \gamma_5 \right) \gamma^\mu ~,
\end{eqnarray}
where \,$2\hat{V}_f^{E}=\hat{L}_f^{E}+\hat{R}_f^{E}$\, and \,$2\hat{A}_f^{E}=\hat{L}_f^{E}-\hat{R}_f^{E}$.\,
From the kinetic term of the $C_k$ gauge bosons, \,$-\frac{1}{4} C_{k\mu \nu} C_k^{\mu \nu}$,\, where
\,$C_k^{\mu \nu} = \partial^\mu C_k^\nu - \partial^\nu C_k^\mu - g_X^{}\epsilon_{krs}^{} C_r^\mu C_s^\nu$,\, with
\,$C_3=Z_{L}\cos\theta-Z_{H}\sin\theta$,\, we derive the Feynman rules
\begin{eqnarray}
& X^\mu X^{\nu\dagger} Z_L^\rho : \;-ig_{X\,}^{}\cos \theta \left[ \bigl(p_X^\rho-p_{X^\dagger}^\rho\bigr) g^{\mu \nu} + \bigl(p_{X^\dagger}^\mu-p_{Z_L}^\mu\bigr) g^{\nu \rho} + \bigl(p_{Z_L}^\nu-p_X^\nu\bigr) g^{\mu \rho} \right] , \\
& X^\mu X^{\nu\dagger}Z_H^\rho :  \; -i g_{X\,}^{}\sin \theta \left[ \bigl(p_X^\rho-p_{X^\dagger}^\rho\bigr) g^{\mu \nu} +\bigl(p_{X^\dagger}^\mu-p_{Z_H}^\mu\bigr) g^{\nu \rho} + \bigl(p_{Z_H}^\nu-p_X^\nu\bigr) g^{\mu \rho} \right] , \\
& X_\mu X_\nu X^\dagger_\rho X^\dagger_\sigma : \; i g_X^2 (g_{\mu \nu}g_{\rho \sigma}-g_{\mu \rho}g_{\nu \sigma}) ~, \\
& X_\mu X_\nu^\dagger Z_{L \rho}Z_{L \sigma} : \; i g_X^2 \cos^2\!\theta\, (g_{\mu \nu}g_{\rho \sigma}-g_{\mu \rho}g_{\nu \sigma}) ~, \\
& X_\mu X_\nu^\dagger Z_{H \rho}Z_{H \sigma} : \; i g_X^2 \sin^2\!\theta\, (g_{\mu \nu}g_{\rho \sigma}-g_{\mu \rho}g_{\nu \sigma}) ~,\\
& X_\mu X_\nu^\dagger Z_{L \rho}Z_{H \sigma} : \; 2i g_X^2 \sin\theta\, \cos \theta\, (g_{\mu \nu}g_{\rho \sigma}-g_{\mu \rho}g_{\nu \sigma}) ~,
\end{eqnarray}
where the momenta are all incoming.


\section{Kinetic mixing between U(1$\bm{)_Y}$ and U(1$\bm{)_{B-L}}$\label{KM}}

In the gauge \,SU(2)$_L\times$U(1)$_Y\times$SU(2)$_X\times$U(1)$_{B-L}$\, sector of our model,
the gauge bosons that may undergo mixing are $W_3$, $\cal B$, $C_3$, and~$E$, respectively.
We can express the Lagrangian for the kinetic and mass terms of these particles after
electroweak symmetry breaking as
\begin{eqnarray}
{\cal L}_G^{} &=& -\mbox{$\frac{1}{4}$}\,W_3^{\alpha\omega}W_{3\alpha\omega}^{}
- \mbox{$\frac{1}{4}$}\,{\cal B}^{\alpha\omega}{\cal B}_{\alpha\omega}^{}
- \mbox{$\frac{1}{4}$}\,C_3^{\alpha\omega}C_{3\alpha\omega}^{}
- \mbox{$\frac{1}{4}$}\,E^{\alpha\omega}E_{\alpha\omega}^{}
- \mbox{$\frac{1}{2}$}\,\sin_{\!}\chi\; {\cal B}^{\alpha\omega}E_{\alpha\omega}^{}
\nonumber \\ && \! +\;
\mbox{$\frac{1}{2}$}\,m_{W\,}^2 W_3^2 + \mbox{$\frac{1}{2}$}\,m_{\cal B\,}^2{\cal B}^2
- m_{W\,}^{}m_{\cal B\,}^{}W_3^\alpha{\cal B}_\alpha^{}
+ \mbox{$\frac{1}{2}$}\,m_{C_3\,}^2 C_3^2 + \mbox{$\frac{1}{2}$}\,m_{E\,}^2 E^2
+ m_{C_3\,}^{} \mu_{E\,}^{} C_3^\alpha E_\alpha^{}
\nonumber \\ &=&
-\mbox{$\frac{1}{4}$}\,G_{\alpha\omega}^{\rm T}\,K\,G^{\alpha\omega} \,+\,
\mbox{$\frac{1}{2}$}\,G_\alpha^{\rm T}\,M_G^2\,G^\alpha ~,
\end{eqnarray}
where \,$f_{\alpha\omega}=\partial_\alpha f_\omega-\partial_\omega f_\alpha$,\,
the $\sin_{\!}\chi$ term describes kinetic mixing between the hypercharge and $B$$-$$L$
gauge bosons, $\cal B$ and $E$, respectively,
\begin{eqnarray} & \displaystyle
m_W^{} \,\,=\,\, \mbox{$\frac{1}{2}$}\,g_{L\,}^{}v_H^{} ~, \hspace{5ex}
m_{\cal B}^{} \,\,=\,\, \mbox{$\frac{1}{2}$}\,g_{Y\,}^{}v_H^{} ~,
& \\ & \displaystyle
m_{C_3}^{} \,\,=\,\, 2 g_{X\,}^{} v_\Phi^{} ~, \hspace{5ex}
m_E^2 \,\,=\,\, 4 g_{B-L\,}^2 v_S^2 + \mu_E^2 ~, \hspace{5ex}
\mu_E^{} \,\,=\,\, 2 g_{B-L\,}^{} v_\Phi^{} ~, ~~~~
& \\ & \displaystyle \hspace{-4ex}
G \,= \left(\!\begin{array}{c} {\cal B}^{\vphantom{|}} \vspace{3pt} \\ W_3^{} \vspace{3pt} \\
E \vspace{3pt} \\ C_3^{} \end{array}\!\right)^{\vphantom{|}} , \hspace{3ex}
K \,= \left(\!\begin{array}{cccc} 1 & 0 & s_\chi^{} & 0 \vspace{3pt} \\ 0 & 1 & 0 & 0
\vspace{3pt} \\ s_\chi^{} & 0 & 1 & 0 \vspace{3pt} \\ 0 & 0 & 0 & 1 \end{array}\!\right) ,
\hspace{3ex}
M_G^2 \,= \left(\!\begin{array}{cccc}
m_{\cal B\,}^{2^{\vphantom{|}}} & -m_{\cal B\,}^{}m_W^{} & 0 & 0 \vspace{3pt} \\
-m_{\cal B\,}^{}m_W^{} & m_W^2 & 0 & 0 \vspace{3pt} \\
0 & 0 & m_E^2 & \mu_{E\,}^{} m_{C_3}^{} \vspace{3pt} \\
0 & 0 & \mu_{E\,}^{} m_{C_3}^{} & m_{C_3}^2  \end{array}\!\right) , &
\end{eqnarray}
with \,$s_\chi^{}=\sin\chi$\, and $g_L^{}$ and $g_Y^{}$ being the SU(2)$_L\times$U(1)$_Y$
gauge couplings, respectively.
In ${\cal L}_G$ we have included the $s_\chi^{}$ term because it does not violate any of
the symmetries in the theory, implying that in general $s_\chi^{}$ can receive both tree- and
loop-level contributions~\cite{Holdom:1985ag,Foot:1991kb}, the latter being due to the SM
fermions carrying both the U(1)$_Y$ and U(1)$_{B-L}$ charges.

It is straightforward to demonstrate that one can convert the kinetic part of ${\cal L}_G^{}$
into the canonical form, \,$-\frac{1}{4}\,\hat G_{\alpha\omega}^{\rm T}\hat G^{\alpha\omega}$,\,
and diagonalize the $M_G^2$ matrix by making the transformation
\begin{eqnarray} \label{mixing_M}
G \,\,=\,\, \tilde T_{\,} O_{\rm w\,}^{} {\cal O}_z^{} \left(\!\begin{array}{c} A \\
Z \\ Z_H^{} \\ Z_L^{} \end{array}\!\right) ,
\end{eqnarray}
where $A$, $Z$, $Z_H$, and $Z_L$ are the mass eigenstates, the photon $A$ staying massless,
\begin{eqnarray} & \displaystyle
\tilde T \,\,=\, \left(\!\begin{array}{cccc} 1 & 0 & -t_\chi^{} & 0 \vspace{3pt} \\
0 & 1 & 0 & 0 \vspace{3pt} \\ 0 & \, 0 \, & \, 1/c_\chi^{} \, & 0 \vspace{3pt} \\
0 & 0 & 0 & 1 \end{array}\!\right) , \hspace{5ex}
O_{\rm w}^{} \,\,=\,
\left(\!\begin{array}{cccc} c_{\rm w}^{} & -s_{\rm w}^{} & 0 & 0 \vspace{3pt} \\
s_{\rm w}^{} & c_{\rm w}^{} & 0 & 0 \vspace{3pt} \\ 0 & 0 & 1 & 0 \vspace{3pt} \\
0 & 0 & 0 & 1 \end{array}\!\right) , \hspace{5ex}
{\cal O}_z^{} \,\,=\, \left(\!\begin{array}{cccc} 1 & 0 & 0 & 0 \vspace{3pt} \\
0 & {\cal O}_{11} & {\cal O}_{12} & {\cal O}_{13} \vspace{3pt} \\
0 & \, {\cal O}_{21} \, & \, {\cal O}_{22} \, & {\cal O}_{23} \vspace{3pt} \\
0 & {\cal O}_{31} & {\cal O}_{32} & {\cal O}_{33} \end{array}\!\right)_{\vphantom{|}} , ~~~~~
& \\ & \displaystyle
c_\chi^{} \,\,=\,\, \cos\chi ~, ~~~~~~~ t_\chi^{} \,\,=\,\, \tan\chi ~, ~~~~~~~
c_{\rm w}^{} \,\,=\,\, \cos\theta_{\rm W}^{} ~, ~~~~~~~
s_{\rm w}^{} \,\,=\,\, \sin\theta_{\rm W}^{} \,\,=\,\,
\frac{c_{\rm w\,}^{}m_{\cal B}^{}}{m_W^{}} ~. &
\end{eqnarray}
The $O_{\rm w}$ and ${\cal O}_z$ matrices are orthogonal, while $\tilde T$ is not.
The elements ${\cal O}_{ij}$ encode the effect of the kinetic mixing, such that in
its absence they are given by \,${\cal O}_{11}=1$,\,
\,${\cal O}_{12}={\cal O}_{21}={\cal O}_{13}={\cal O}_{31}=0$,\,
\,${\cal O}_{22}={\cal O}_{33}=\cos\theta$,\, and
\,${\cal O}_{23}=-{\cal O}_{32}=\sin\theta$,\,
which is the case treated in the main text.

It is also simple to see that the presence of kinetic mixing, \,$s_\chi^{}\neq0$,\, affects
all of the couplings of $Z$ and $Z_{L,H}$ to fermions.
Since the fermions do not couple directly to the $C_3$ gauge boson, one can write
the Lagrangian for their interactions with the $W_3^{}$, $\cal B$, and $E$ bosons in
terms of the physical states $A$, $Z$, and $Z_{H,L}$ as
\begin{eqnarray} \label{correction-coupling}
{\cal L}' &=&
-g_{L\,}^{}J_3^\lambda\,W_{3\lambda}^{}\,-\,g_{Y\,}^{}J_Y^\lambda\,{\cal B}_\lambda^{}
\,-\,g_{B-L\,}^{}J_{B-L}^\lambda\,E_\lambda^{}
\nonumber \\ &=&
-e_{\,} J_{\rm em\,}^\lambda A_\lambda^{} \,-\, \Biggl[
\bigl({\cal O}_{11}^{}\,+\,{\cal O}_{21\,}^{}t_{\chi\,}s_{\rm w}^{}\bigr)\,
\hat g_{Z\,}^{}\hat J_Z^\lambda
\,+\, \frac{{\cal O}_{21}^{}}{c_\chi^{}}\,g_{B-L\,}^{}J_{B-L}^\lambda
\,-\, {\cal O}_{21\,}^{}t_{\chi\,}^{}c_{\rm w\,}^{}e_{\,}J_{\rm em}^\lambda \Biggr] Z_\lambda^{}
\nonumber \\ && \! -\;
\Biggl[ \frac{{\cal O}_{22}^{}}{c_\chi^{}}\, g_{B-L\,}^{}J_{B-L}^\lambda +
\bigl({\cal O}_{12}^{}\,+\,{\cal O}_{22\,}^{}t_{\chi\,}s_{\rm w}^{}\bigr)\,
\hat g_{Z\,}^{}\hat J_Z^\lambda
\,-\, {\cal O}_{22\,}^{}t_{\chi\,}^{}c_{\rm w\,}^{}e_{\,}J_{\rm em}^\lambda \Biggr] Z_{H\lambda}^{}
\nonumber \\ && \! -\;
\Biggl[ \frac{{\cal O}_{23}^{}}{c_\chi^{}}\,g_{B-L\,}^{}J_{B-L}^\lambda +
\bigl({\cal O}_{13}^{}\,+\,{\cal O}_{23\,}^{}t_{\chi\,}s_{\rm w}^{}\bigr)\,
\hat g_{Z\,}^{}\hat J_Z^\lambda
\,-\, {\cal O}_{23\,}^{}t_{\chi\,}^{}c_{\rm w\,}^{}e_{\,}J_{\rm em}^\lambda \Biggr] Z_{L\lambda}^{} ~,
\end{eqnarray}
where $J_{3,Y,B-L}$ are the currents coupled to the respective fields and we have used
the relations
\begin{eqnarray}
e \,\,=\,\, g_{L\,}^{}s_{\rm w}^{} \,\,=\,\, g_{Y\,}^{}c_{\rm w}^{} ~, \hspace{5ex}
J_{\rm em}^{} \,\,=\,\, J_3^{} \,+\, J_Y^{} ~, \hspace{5ex}
\hat g_{Z\,}^{}\hat J_Z^{} \,\,=\,\,
c_{\rm w\,}^{}g_{L\,}^{}J_3^{}\,-\,s_{\rm w\,}^{}g_{Y\,}^{}J_Y^{} ~.
\end{eqnarray}

From the previous paragraphs, one can infer that the $\chi$-dependent new terms translate
into modifications to $Z$-pole observables and the \,$e^+ e^-\to f\bar f$\, cross-sections,
as well as the $Z$ and $Z_{H,L}$ masses.
Consequently, such contributions must respect the pertinent experimental restrictions.
After imposing them, we find that for the ranges of the new gauge couplings and $Z_L$ mass
satisfying the relic data requirement the kinetic-mixing effects are unimportant on
the $Z$ and $Z_L$ masses, but for $\sin\chi$ not much less than~1 they could enlarge
the $Z_H$ mass substantially compared to that in the \,$\chi=0$\, case.
Specifically, the increase in $m_{Z_H}$ would be mild, no more than about 15\%,
if~\,$|\sin\chi|\lesssim0.5$.\,

To see if such mixing size is reasonable, we consider the two-point polarization diagram
for the $\cal B$ and $E$ gauge bosons with fermions in the loop.
Accordingly, we estimate the kinetic mixing parameter to be~\cite{Holdom:1985ag,Dienes:1996zr}
\begin{equation}
\label{mixing-angle}
\sin \chi \,\,\simeq\,\,  -\sum_f \frac{g_{B-L\,}^{}g_Y^{}}{24\pi^2}\, (B_f-L_f) Y_f\,
\ln \frac{\bigl|q^2\bigr|}{\Lambda^2} ~,
\end{equation}
where the sum is over the SM chiral fermions, $B_f-L_f$ and $Y_f$ denote the $B-L$ and
hypercharge values for fermion $f$, respectively, $q$ is the renormalization scale which we
take to be the U(1)$_{B-L}$ breaking scale, of order 1\,TeV or greater, and
we have applied the renormalization condition that at some higher scale $\Lambda$ the sum of
the loop and counterterm contributions vanishes.
We note that one could evaluate $\sin\chi$ more precisely using the renormalization group
equation to resum the large logarithms~\cite{Dienes:1996zr, Babu:1996vt}, but the difference
would amount to only a few percent for our scales of interest and
therefore can be ignored.
Thus, since \,$\sum_f (B_f-L_f) Y_f=8$,\, taking \,$\Lambda^2\sim10^{6\,}\bigl|q^2\bigr|$\,
in Eq.\,(\ref{mixing-angle}) we get \,$\sin\chi\sim 0.16\,g_{B-L}^{}$.\,
For the viable $g_{B-L}^{}$ values we have obtained, this result is compatible with
the $|\sin\chi|$ number quoted in the last paragraph.

We remark that this is also consistent with the findings of a detailed analysis in
Ref.\,\cite{Williams:2011qb} on the phenomenological constraints on a new massive Abelian gauge boson.
The effects of such a particle can be compared to those of $Z_H$ which contains mostly its
U(1)$_{B-L}$ component~$E$ and has a~mass of ${\cal O}(1$-10)\,TeV in our study.
For a new massive Abelian gauge boson in this mass range, the results of
Ref.\,\cite{Williams:2011qb} imply \,$|\sin\chi|\lesssim 0.6\,$-1.\,
They also do not lead to additional restraints on the corresponding viable values of $g_{B-L}^{}$.




\begin{thebibliography}{99}

\bibitem{pdg}
  J.~Beringer {\it et al.}  [Particle Data Group Collaboration],
  Phys.\ Rev.\ D {\bf 86}, 010001 (2012).

\bibitem{CNT}
  C.W.~Chiang, T.~Nomura, and J.~Tandean,
  Phys.\  Rev.\ D {\bf 87}, 073004 (2013)
  [arXiv:1205.6416 [hep-ph]].

\bibitem{Krauss:1988zc}
  L.M.~Krauss and F.~Wilczek,
  Phys.\ Rev.\ Lett.\  {\bf 62}, 1221 (1989).

\bibitem{seesaw}
  P.~Minkowski,
  Phys.\ Lett.\  B {\bf 67}, 421 (1977);
T.~Yanagida, in {\it Proceedings of the Workshop on the Unified Theory and the Baryon Number in
the Universe}, edited by O.~Sawada and A.~Sugamoto (KEK, Tsukuba, 1979), p.~95;
  Prog.\ Theor.\ Phys.\  {\bf 64}, 1103 (1980);
M.~Gell-Mann, P.~Ramond, and R.~Slansky,
in {\it Supergravity},
edited by P.~van Nieuwenhuizen and D.~Freedman (North-Holland, Amsterdam, 1979), p.~315;
S.L.~Glashow, in {\it Proceedings of the 1979 Cargese Summer Institute on Quarks and Leptons},
edited by M.~Levy {\it et al}. (Plenum Press, New York, 1980), p. 687;
  R.N.~Mohapatra and G.~Senjanovic,
  Phys.\ Rev.\ Lett.\  {\bf 44}, 912 (1980);
   J.~Schechter and  J.W.F.~Valle,
    Phys.\  Rev.\ D {\bf 22}, 2227 (1980);
    Phys.\  Rev.\ D {\bf 25}, 774 (1982).

\bibitem{vectordm}
  T.~Hambye,
  JHEP {\bf 0901}, 028 (2009)
  [arXiv:0811.0172 [hep-ph]];
  T.~Hambye and M.H.G.~Tytgat,
  Phys.\ Lett.\ B {\bf 683}, 39 (2010)
  [arXiv:0907.1007 [hep-ph]];
  F.~Chen, J.M.~Cline, and A.R.~Frey,
  Phys.\ Rev.\ D {\bf 80}, 083516 (2009)
  [arXiv:0907.4746 [hep-ph]];
  J.L.~Diaz-Cruz and E.~Ma,
  Phys.\ Lett.\ B {\bf 695}, 264 (2011)
  [arXiv:1007.2631 [hep-ph]];
  S.~Bhattacharya, J.L.~Diaz-Cruz, E.~Ma, and D.~Wegman,
  Phys.\ Rev.\ D {\bf 85}, 055008 (2012)
  [arXiv:1107.2093 [hep-ph]];
  O.~Lebedev, H.M.~Lee, and Y.~Mambrini,
  Phys.\ Lett.\ B {\bf 707}, 570 (2012)
  [arXiv:1111.4482 [hep-ph]];
  Y.~Farzan and A.R.~Akbarieh,
  JCAP {\bf 1210}, 026 (2012)
  [arXiv:1207.4272 [hep-ph]];
  T.~Abe, M.~Kakizaki, S.~Matsumoto, and O.~Seto,
  Phys.\ Lett.\ B {\bf 713}, 211 (2012)
  [arXiv:1202.5902 [hep-ph]];
  F.~D'Eramo, M.~McCullough, and J.~Thaler,
  JCAP {\bf 1304}, 030 (2013)
  [arXiv:1210.7817 [hep-ph]].

\bibitem{Alcaraz:2006mx}
ALEPH Collaboration, DELPHI Collaboration, L3 Collaboration, OPAL Collaboration, and the LEP
Electroweak Working Group (The LEP Collaborations),
  arXiv:hep-ex/0612034.

\bibitem{Zprime}
  C.W.~Chiang, Y.F.~Lin, and J.~Tandean,
  JHEP {\bf 1111}, 083 (2011)  [arXiv:1108.3969 [hep-ph]].

\bibitem{dy@lhc}
CMS Collaboration,
Report No. CMS-PAS-EWK-11-007, April 2012.

\bibitem{CW}
  C.W.~Chiang, N.D.~Christensen, G.J.~Ding, and T.~Han,
  Phys.\ Rev.\  D {\bf 85}, 015023 (2012)  [arXiv:1107.5830 [hep-ph]].

\bibitem{Ref:CalcHEP}
  A.~Pukhov {\it et al.},
  arXiv:hep-ph/9908288.

\bibitem{Kolb:1990vq}
E.W. Kolb and M. Turner, {\it The Early Universe} (Westview Press, Boulder, 1990);
  K.~Griest and D.~Seckel,
  Phys.\ Rev.\  D {\bf 43}, 3191 (1991).

\bibitem{TA}
  P.~Gondolo and G.~Gelmini,
  Nucl.\ Phys.\ B {\bf 360}, 145 (1991).

\bibitem{planck}
  P.A.R.~Ade {\it et al.}  [Planck Collaboration],
  arXiv:1303.5062 [astro-ph.CO].

\bibitem{NucleonMatrixElement}
  D.B.~Kaplan and A.~Manohar,
  Nucl.\ Phys.\  B {\bf 310}, 527 (1988).

\bibitem{XENON10}
  J.~Angle {\it et al.}  [XENON10 Collaboration],
Phys.\ Rev.\ Lett.\  {\bf 107}, 051301 (2011)  [arXiv:1104.3088 [astro-ph.CO]];
{\bf 110}, 249901(E) (2013).

\bibitem{Frandsen:2013cna}
M.T.~Frandsen, F.~Kahlhoefer, C.~McCabe, S.~Sarkar, and K.~Schmidt-Hoberg,
  JCAP {\bf 1307}, 023 (2013)  [arXiv:1304.6066 [hep-ph]].

\bibitem{XENON100}
  E.~Aprile {\it et al.}  [XENON100 Collaboration],
  Phys.\ Rev.\ Lett.\  {\bf 109}, 181301 (2012)
  [arXiv:1207.5988 [astro-ph.CO]].

\bibitem{CDMS-Ge}
  Z.~Ahmed {\it et al.}  [CDMS-II Collaboration],
  Science {\bf 327}, 1619 (2010)
  [arXiv:0912.3592 [astro-ph.CO]];
  D.S.~Akerib {\it et al.}  [CDMS Collaboration],
  Phys.\ Rev.\ D {\bf 82}, 122004 (2010)
  [arXiv:1010.4290 [astro-ph.CO]];
  Z.~Ahmed {\it et al.}  [CDMS-II Collaboration],
  Phys.\ Rev.\ Lett.\  {\bf 106}, 131302 (2011)
  [arXiv:1011.2482 [astro-ph.CO]].

\bibitem{CDMS-Si}
  R.~Agnese {\it et al.}  [CDMS Collaboration],
  Phys.\ Rev.\ D {\bf 88}, 031104 (2013)
  [arXiv:1304.3706 [astro-ph.CO]].

\bibitem{Akerib:2013tjd}
  D.S.~Akerib {\it et al.}  [LUX Collaboration],
  arXiv:1310.8214 [astro-ph.CO].

\bibitem{cogent}
  C.E.~Aalseth {\it et al.}  [CoGeNT Collaboration],
  arXiv:1208.5737 [astro-ph.CO];
  Phys.\ Rev.\ Lett.\  {\bf 106}, 131301 (2011)
  [arXiv:1002.4703 [astro-ph.CO]];
  Phys.\ Rev.\ Lett.\  {\bf 107}, 141301 (2011)
  [arXiv:1106.0650 [astro-ph.CO]].

\bibitem{DAMA}
  C.~Savage, G.~Gelmini, P.~Gondolo, and K.~Freese,
  JCAP {\bf 0904}, 010 (2009)
  [arXiv:0808.3607 [astro-ph]];
  R.~Bernabei {\it et al.}  [DAMA and LIBRA Collaborations],
  Eur.\ Phys.\ J.\ C {\bf 67}, 39 (2010)  [arXiv:1002.1028 [astro-ph.GA]].

\bibitem{CRESST}
  G.~Angloher {\it et al.} [CRESST Collaboration],
  Eur. \ Phys. \ J. \ C {\bf 72}, 1971 (2012)
  arXiv:1109.0702 [astro-ph.CO].

\bibitem{CDMSII}
  R.~Agnese {\it et al.}  [CDMS Collaboration],
arXiv:1304.4279 [hep-ex].

\bibitem{xenon1t}
E. Aprile, 
  arXiv:1206.6288 [astro-ph.IM].

\bibitem{susylimit}
  G.~Aad {\it et al.}  [ATLAS Collaboration],
  Phys.\ Rev.\ D {\bf 87}, 012008 (2013)
  [arXiv:1208.0949 [hep-ex]].

\bibitem{Steigman:2012nb}
G.~Steigman, B.~Dasgupta, and J.F.~Beacom,
Phys.\ Rev.\ D {\bf 86}, 023506 (2012)
[arXiv:1204.3622 [hep-ph]].

\bibitem{Holdom:1985ag}
B.~Holdom,
Phys.\ Lett.\ B {\bf 166}, 196 (1986)

\bibitem{Foot:1991kb}
F.~del Aguila, G.D.~Coughlan, and M.~Quiros,
  Nucl.\ Phys.\ B {\bf 307}, 633 (1988)
  [Erratum-ibid.\ B {\bf 312}, 751 (1989)];
R.~Foot and X.G.~He,
  Phys.\ Lett.\ B {\bf 267}, 509 (1991);
R.~Foot, X.G.~He, H.~Lew, and R.R.~Volkas,
  Phys.\ Rev.\ D {\bf 50}, 4571 (1994)
  [hep-ph/9401250].

\bibitem{Dienes:1996zr}
K.R.~Dienes, C.~Kolda, and J.~March-Russell,
  Nucl.\ Phys.\  B {\bf 492}, 104 (1997).

\bibitem{Babu:1996vt}
K.S.~Babu, C.~Kolda, and J.~March-Russell,
Phys.\ Rev.\ D {\bf 54}, 4635 (1996),

\bibitem{Williams:2011qb}
M.~Williams, C.P.~Burgess, A.~Maharana, and F.~Quevedo,
  JHEP {\bf 1108}, 106 (2011)  [arXiv:1103.4556 [hep-ph]].


\end{thebibliography}
\end{document}